\documentclass[fleqn,usenatbib]{mnras}

\usepackage{newtxtext,newtxmath}

\usepackage[T1]{fontenc}

\DeclareRobustCommand{\VAN}[3]{#2}
\let\VANthebibliography\thebibliography
\def\thebibliography{\DeclareRobustCommand{\VAN}[3]{##3}\VANthebibliography}

\usepackage{graphicx}	
\usepackage{amsmath}	
\usepackage{tabularx}

\title{DETECTION OF TWO NEW RRATS AT 111 MHz}

\author[IPS and pulse statistics]
{ V. A. Samodurov,$^{a,b}$ 
S. A. Tyul'bashev$^{b}$\thanks{E-mail: serg@prao.ru}
 M. O. Toropov $^{c}$, 
	A. V. Dolgushev,$^{d}$,
	V. V. Oreshko$^{b}$,
	S. V. Logvinenko$^{b}$\\
$^{a}$ National Research University Higher School of Economics, Moscow, Russia\\
$^{b}$ Lebedev Physical Institute, Astro Space Center, Pushchino Radio Astronomical Observatory, Pushchino, Russia\\
$^{c}$ TEK Inform, Moscow, Russia\\
$^{d}$ Yandex, Moscow, Russia\\
}

\date{ }

\pubyear{ }

\begin{document}
	\label{firstpage}
	\pagerange{\pageref{firstpage}--\pageref{lastpage}}
	\maketitle
	
	\begin{abstract}
A search for pulse signals in a area with declinations of $+52\degr <\delta <+55\degr$ was carried out on the LPA LPI radio telescope. When processing ten months of observations recorded in six frequency channels with a channel width of 415 kHz and a total bandwidth of 2.5 MHz, 22 thousand events were found with a pronounced dispersion delay of signals over frequency channels, i.e. having signs of pulsar pulses. It turned out that the found pulses belong to four known pulsars and two new rotating radio transients (RRATs). An additional pulse search conducted in 32-channel data with a channel width of 78 kHz revealed 8 pulses for the transient J0249+52 and 7 pulses for the transient J0744+55. Periodic radiation of transients was not detected. The analysis of observations shows that the found RRATs are most likely pulsars with nullings, where the proportion of nulling is greater than 99.9\%.
	\end{abstract}
	
	
	
	
	\section{Introduction}
    \label{intro}
In 2014, the radio telescope Large Phased Array (LPA) of the Lebedev Physical Institute of the Russian Academy of Sciences (LPI) after the modernization of the antenna, during which the fluctuation sensitivity of the radio telescope increased 2-3 times, and the number of simultaneously observed beams increased to 128, regular monitoring observations were started. These observations are used for the study of interplanetary plasma (\citeauthor{Shishov2016}, \citeyear{Shishov2016}), the search for pulsars and transients (\citeauthor{Tyul'bashev2018}, \citeyear{Tyul'bashev2018}), the study of variability active galactic nuclei (\citeauthor{Tyul'bashev2019}, \citeyear{Tyul'bashev2019}). The increased sensitivity of the radio telescope makes it possible to register individual pulses of more than 150 pulsars (https://bsa-analytics.prao.ru/en/). Some of these pulsars, representing irregularly observed pulse signals, were detected for the first time at the LPA LPI and belong to a sample of rotating radio transients (RRATs).

The optimal search for pulsed dispersed signals (pulses arriving first at a high and then at a low frequency), taking into account their scattering and scintillation in the interstellar medium, was considered in the work of \citeauthor{Cordes2003}, \citeyear{Cordes2003}. The use of the proposed method of searching for pulses in an application to archival data obtained on a 64-meter radio telescope (Parkes, Australia) led in 2006 to the discovery of 11 pulsars with special properties (\citeauthor{McLaughlin2006}, \citeyear{McLaughlin2006}) and in 2007 to the discovery of pulses of extragalactic nature (\citeauthor{Lorimer2007}, \citeyear{Lorimer2007}). If ordinary pulsars emit pulses at every or almost every revolution, and their search can be carried out in a standard way using power spectra or using periodograms, then many periods can pass between the observed consecutive pulses without pulse emission in RRAT. Therefore, when searching by standard methods, such pulsars are not detected.

The total number of known RRAT is small. In the ATNF database (https://www.atnf.csiro.au/research/pulsar/psrcat/; \citeauthor{Manchester2005}, \citeyear{Manchester2005}) and RRATalog (http://astro.phys.wvu.edu/rratalog/) there are approximately 100-150 rotating radio transients. More recently, work has appeared to search for RRAT on the FAST radio telescope, and it reports the detection of 76 new transients (\citeauthor{Zhou2023}, \citeyear{Zhou2023}). Together with paper \citeauthor{Zhou2023}, \citeyear{Zhou2023}, the number of known RRAT exceeded 200. At the same time, estimates show that the expected number of RRAT can be twice as large as the number of ordinary pulsars (\citeauthor{Keane2008}, \citeyear{Keane2008}). That is, an insignificant part of the RRATs has been detected. Almost all RRATs are open on several radio telescopes with high instantaneous sensitivity: the Parkes 64-m telescope (Australia) (\citeauthor{McLaughlin2006}, \citeyear{McLaughlin2006}); the Arecibo telescope, with a diameter of 300 m (Puerto Rico) (\citeauthor{Deneva2013}, \citeyear{Deneva2013}); Green Bank Telescope (GBT) with a diameter of 100 m (USA) (\citeauthor{Stovall2014}, \citeyear{Stovall2014}); LPA LPI telescope, which is a $200\times 400$ m antenna array (Russia) (\citeauthor{Tyul'bashev2018}, \citeyear{Tyul'bashev2018}); FAST telescope with a diameter of 500 m (China) (\citeauthor{Zhou2023}, \citeyear{Zhou2023}); CHIME telescope with a size of $80\times 100$~m (Canada) (\citeauthor{Dong2022}, \citeyear{Dong2022}).

The appearance of RRAT pulses is unpredictable. The typical time between two consecutive pulses can be from minutes to hours (\citeauthor{McLaughlin2006}, \citeyear{McLaughlin2006}), but can reach tens of hours (\citeauthor{Logvinenko2020}, \citeyear{Logvinenko2020}). In \citeauthor{Smirnova2022}, \citeyear{Smirnova2022}, two RRATs (J1132+25, J1336+33) were noted, in which not a single pulse was recorded in more than 250 observation sessions that took place once a day and had a duration of about 3.5 minutes. Thus, to search for RRAT, in addition to an antenna with high fluctuation sensitivity, it is necessary to have long-term of observations.

During the observations, pulse signals are constantly recorded at the LPA LPI. A special study to check the quality of data obtained during monitoring showed that about a fifth of these pulse signals are associated with pulsars, and all the rest are interference (\citeauthor{Samodurov2022}, \citeyear{Samodurov2022}). The statistical approach used to check the quality of monitoring data made it possible to detect from several to several thousand pulses of six known pulsars, as well as to discover 4 new RRATs in the data recorded on the new recorder (\citeauthor{Samodurov2022}, \citeyear{Samodurov2022}).

In this paper, short duration signals detected in raw data are investigated. We consider the statistics of detection of pulsed radiation sources in a new area included in the monitoring program for the search for pulsars in the test mode in the fall of 2021, and the identification of these pulsed sources with interference and real signals of extraterrestrial origin.

\begin{figure*}
	\includegraphics{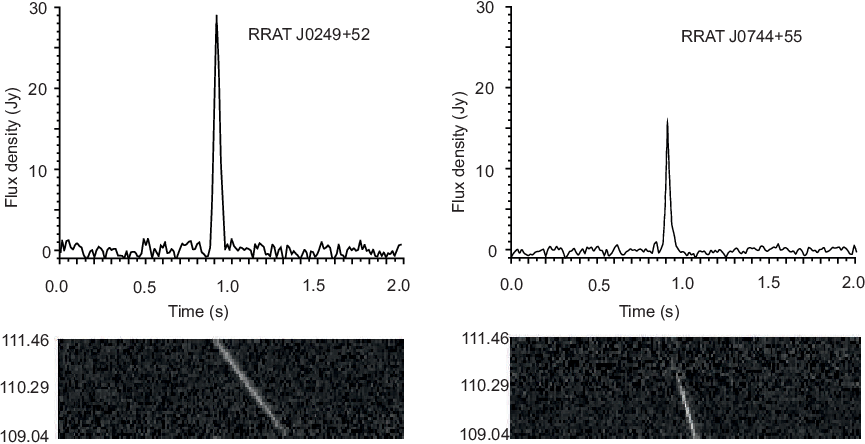}
	\caption{Top panel: profiles of the strongest pulses of the found transients. The vertical axis is the flux density in janskys. Bottom	panel: dynamic spectra of these pulses. The vertical axis represents the frequencies of several channels. The time intervals on the	horizontal axis of the pulse profile and the corresponding dynamic spectrum coincide.}
	\label{Fig1}
\end{figure*}

\section{Observations}

The LPA LPI radio telescope is a meridian instrument. Any source in the sky can be observed once a day, and the typical duration of the observation session it takes 3-4 minutes. During the modernization of the antenna, it was possible to implement a scheme that allows 4 independent radio telescopes to be launched on the basis of one antenna field. 
At the moment, two radio telescopes have been put into operation. 
One of them (LPA1) is used for standard observations of pulsars. 
Its 512 beams cover declinations in the range $-15\degr < \delta < +87\degr$ with overlapping beams at the level of 0.8. 
The second telescope (LPA3) is 128-beam and covers declinations of $-9\degr < \delta <+55\degr$ when the rays overlap in level 0.4.
If the coordinates of the source coincide with the coordinates of the beam, then the source is not observed in the beams above and below.

In the period 2013-2014, 96 beams overlapping declinations $-9\degr < \delta <+42\degr$ were connected to two recorders of the 128-beam radio telescope. At the end of 2020, a new recorder was created, to which 24 beams were connected in test mode, overlapping declinations of $+42\degr < \delta <+52\degr$. The quality of observations in these 24 beams was analyzed, and the results of the analysis were published in \citeauthor{Samodurov2022}, \citeyear{Samodurov2022}. In addition to evaluating the quality of data in the newly connected beams, four new RRATs were also detected. In 2021, the last 8 beams were put into operation, overlapping declinations of $+52\degr < \delta <+55\degr$. Thus, all available beams are used on LPA3 (128 = 48 + 48 + (24 + 8)), to which three blocks with recorders are attached. The testing period of the equipment as a whole ended on 20 October 2021, after which test observations were started in the round-the-clock monitoring mode. This paper presents the first results of data analysis in the period from 21 October 2021 to 31 August 2022.

The characteristics of LPA3 are as follows: the central receiving frequency is 110.25 MHz; the received band is 2.5 MHz; the effective antenna area is about 45,000 sq.m. Raw data is synchronously recorded in two time-frequency resolutions. In data with low time-frequency resolution, the reception band is divided into 6 frequency channels with a width of 415 kHz. The sampling of the point is 100 ms. These data are used in the Space Weather project (\citeauthor{Shishov2016}, \citeyear{Shishov2016}). Interference (pulse signals) are analyzed based on these data. Data with high time-frequency resolution is recorded in 32-channel mode with a channel width of 78 kHz. The sampling is 12.5 ms. 32-channel data is used when additional verification of the found pulses is necessary. Data is recorded at hourly segments. Immediately after the end of the recording, the recording of the next hour of observations starts.

To equalize the gain in the frequency channels, a noise signal of a known temperature is used, supplied to a distributed amplification system. It is recorded as an OFF-ON-OFF (calibration step), where the OFF mode means that there is no calibration signal when all intermediate amplifiers are turned off. In this case, the noise in the antenna paths corresponding to the ambient temperature is prescribed. The ON mode corresponds to the activation of the calibration signal (temperature 2400 K) when the dipole lines are switched off. For more information about working with the calibration step, see the paper \citeauthor{Tyul'bashev2019}, \citeyear{Tyul'bashev2019}.

\section{Processing of data}

\begin{table*}[]
	\caption{Characteristics of the found RRAT}
	\begin{tabular}{|c|c|c|c|c|c|c|c|}
		\hline
		Name & $\alpha_{2000}$ (h,m,s) & $\delta_{2000}$  ($\degr, ^\prime $) & $DM$ (pc/cm$^3$) & $W_{0.5}$ (ms) & $S_{peak}$ (Jy) & $N_1/N_2$ &  $n$ ($h^{-1}$)\\ \hline
		J0249+52 & 02 49 00$\pm$90 & 52 46$\pm$15 & 27.5$\pm$1.5 & 37-72 & 16.5-29.0 & 8/8 & 0.44\\
		J0744+55 & 07 44 45$\pm$90 & 55 05$\pm$10 & 10.5$\pm$1.5 & 20-42 &  8.8-15.5 & 7/7 &  0.38\\
		\hline
	\end{tabular}
\end{table*}

Previously, the gain in the frequency channels is equalized using a calibration step recorded 6 times a day. Then the data is divided into ten-second segments. For each segment and for each channel, the following are estimated: minimum and maximum intensity values in antenna degrees after calibration by step, median intensity value, rms deviations of noise. For each segment, the date and hour of observations in Moscow time and the beginning of the ten-second segment under study in sidereal time are also stored. The information stored in the database is ten times smaller in volume than the volume of the original raw data. It allows you to detect the level of interference at any selected time interval, as well as to investigate individual pulse interference (events).

Data processing is described in detail in \citeauthor{Samodurov2022}, \citeyear{Samodurov2022}, \citeauthor{Samodurov2017}, \citeyear{Samodurov2017}. Here we also note that the stored coordinate of the observed maximum within a ten-second interval allows you to bind the maxima for each frequency channel in time, and thereby allows you to roughly determine the dispersion delay of the signal in frequency during processing. Since we know the rms deviation of the noise inside the channel, we can search for pulse signals at a given level of S/N. The signal-to-noise ratio is defined as $S/N=A/\sigma_{noise}$, where - $A$ is the amplitude of the signal after subtracting the baseline (background signal), and $\sigma_{noise}$ is the standard deviations at 10 s interval. If in the studied segment for the found pulse the S/N was greater than 5 in at least three frequency channels out of six, it is considered that a candidate for transients has been found. The RRAT candidate is additionally checked against data with high time-frequency resolution.

\section{Results}

In the time span from 21 October 2021 to 31 August 2022, after removing data omissions, 7496 file-hours (312.33 sidereal days) were analyzed. In total, 2.5 million pulses were detected during the processing of 6-channel data. Many impulses were simultaneously manifested in several beams. Such cases were automatically combined into one event. A total of 160504 interrelated events were found.

The distribution of the number of simultaneous events across different beams is similar to the distribution that was in the early analysis of 24 beams (\citeauthor{Samodurov2022}, \citeyear{Samodurov2022}). Most often, the pulse is observed either in one beam or in all beams at the same time. Real pulsar pulses should be observed in one or two adjacent beams. This, of course, does not exclude interference, which can also fall into only one beam. The pulses observed in all beams are, in the vast majority of cases, interference. Individual very powerful pulsar pulses can also be observed in many beams, appearing in the side lobes of the LPA LPI. Other sources of generation of this type of pulses are also possible, but a separate work will be devoted to this.

Analyzing the pulses detected in one or two beams, we found that approximately 22,000 pulses, or 13.9\% of their total number, have a pronounced dispersion delay, i.e. similar to pulsar pulses. The search for new pulsars by their individual dispersed pulses was carried out only for 8 rays out of 32 ($+52\degr<\delta<+55\degr$), since the search for 24 rays was done in an earlier work (\citeauthor{Samodurov2022}, \citeyear{Samodurov2022}).

As in \citeauthor{Samodurov2022}, \citeyear{Samodurov2022}, the area under study was divided into clusters of 2 min. length. For each cluster, the number of pulses found, the coordinates of the right ascension and declination, the average $DM$ of the pulses found, the beam number, the Julian date (MJD), time (UT), sidereal time, the observed S/N in frequency channels, lists of beam numbers with similar pulses were checked. In total, the processing program identified 45 clusters out of 5760 possible clusters that fall on the area under study. Each cluster has a right ascension coordinate determined by a two-minute segment, and a declination coordinate determined by the beam number LPA3. There is at least one "pulsar" event in the allocated 45 clusters. The coordinates of the clusters are determined, and therefore it is possible to identify with the ATNF catalog.

Known strong pulsars can occupy several neighboring clusters at once, both in right ascension and declination. They are easily identified in ATNF, and removed from the most distant analysis. In total, four known pulsars were detected during the analysis: B0329+54 ($P$ = 0.7145c; $DM$ = 26.7 pc/cm$^3$); B0343+53 ($P$ = 1.9344s; $DM$ = 67.3 pc/cm$^3$); B1508+55 ($P$ = 0.7396c; $DM$ = 19.6 pc/cm$^3$) and B2021+51 ($P$ = 0.5291 c; $DM$ = 22.5 pc/cm$^3$). For these pulsars, from one (B0343+53) to more than 10 thousand (B0329+54) pulses were detected. Pulses of pulsars B0329+54 and B1508+55 were also observed in the side lobes.

In addition to the known pulsars, pulses belonging to two new RRAT (J0249+52; J0744+55) were detected in the recordings. Fig.~1 shows their profiles and dynamic spectra. For J0744+55, a profile with a double peak is visible in part of the pulses. Most likely, subpulses are observed in this transient (see Fig.~2). The distance between the maxima in the profile is 25-35 ms.

\begin{figure}
	\includegraphics[width=0.9\columnwidth]{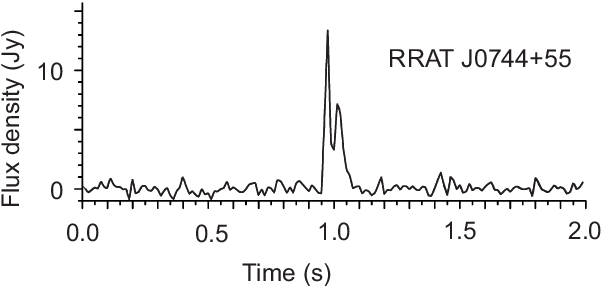}
 	\caption{J0745+55 transient profile showing a double peak.}
	\label{Fig2}
\end{figure}

Table~1 provides information on the found RRAT. Columns 1-3 contain the name of the transient, the coordinates of the source in right ascension and declination. The right ascension was determined as the median value of the detected pulses, and the error, due to the small number of detected pulses, was determined as the size of the half-power LPA beam. The pulses found at J0249+52 are visible in one beam and are not visible in the beams above and below, therefore, the declination coordinate was determined as the declination of the beam, and the accuracy of the coordinate estimate as half the declination distance between the beams. The pulses J0744+55 are visible in two adjacent beams, and the coordinate was determined as the average of the coordinates of the beams in declination. Columns 4-6 show estimates of the $DM$, half-width of the profile ($W_{0.5}$), peak flux densities ($S_{peak}$) of the weakest, and, through a slash ("/"), the strongest of the pulses found. Since a calibration step of a known temperature is prescribed 6 times a day in all frequency channels, the data in the channels are calibrated using a step. Thus, the height of the found pulse is known in units of temperature. Recording is conducted around the clock. In addition to calibration steps, there are also discrete sources with a known flux density in records. Therefore, it is possible to recalculate the observed peak flux densities from temperature units to Jy ones. The estimates given are lower estimates of the flux density. The exact coordinate of the transient is not known either by direct ascension or declination. Therefore, it is impossible to make corrections that take into account the possible hit of the pulse on the edge of the radiation pattern of the LPA and the possible non-coincidence of the coordinates of the beam of the LPA and the coordinates of the transient declination. As a result, the estimate of the $S_{peak}$ can be underestimated up to 2 times. Column 7 indicates how many pulses are detected ($N_1$) from data with high time-frequency resolution. The slash "/" indicates the number of days of pulse detection ($N_2$). Column 8 shows the frequency of occurrence of pulses ($N_1$) having a $S/N > 10$ ($S_{peak} > 3.5-4$~Jy) in 32-channel data for one hour of observations. When receiving the estimate, it was assumed that the pulses most likely appear in the central part of the LPA3 radiation pattern, which is approximately equal to 3.5 minutes in half power. Total accumulated (312 sessions $\times$ 3.5 min) /60 min = 18.2 hours of observations in the direction of each transient.

Approximately the same number of pulses were detected in both RRAT. However, if the transient J0744+55 pulses appeared approximately uniformly throughout the entire observation period, then the transient J0249+52 had all its 8 pulses registered for four consecutive months.

Since RRAT are pulsars, we tried to find their periodic radiation. In the Pushchino multibeam pulsar search (PUMPS) (\citeauthor{Tyul'bashev2022}, \citeyear{Tyul'bashev2022}), a area with declinations of $+52\degr < \delta < +55\degr$ has not been studied before. Assuming that the found RRAT can be ordinary second pulsars, we conducted a standard search using Fourier power spectra. The days with the best quality of the noise track were selected for the search. Out of a total of 312 days of observations, a third of the records were discarded. For the remaining days, using the fast Fourier transform, power spectra were obtained, which were added up. When adding up the spectra, an increase in the $S/N$ harmonics should be observed if there are periodic signals in the studied directions. In the summed power spectra, no details were found in the power spectra at the level $S/N > 5$ and at periods $P<2$~c. In early studies, it was shown that when summing the spectra, the observed growth of the $S/N$ harmonics is less than the root of the number of stacked sessions (\citeauthor{Tyul'bashev2022}, \citeyear{Tyul'bashev2022}). Estimates show that the increase in sensitivity when adding power spectra will be about 10 times, and not 15 times, as one might expect. Taking into account the background temperature in the direction of the transients found, and assuming periods of transients $P < 2$~c, it is possible to give an upper estimate for the expected $S_{peak}$ with regular pulsar radiation: $S_{peak} < 0.3$~Jy (J0249+52), $S_{peak} < 0.2$~Jy (J0744+55). Since the periods of the found RRAT were not found, estimates of the integral flux density could not be obtained.

In addition to using power spectra, an estimate of the RRAT period can be obtained based on the observed time interval between pulses by selecting the largest common divisor (a time interval that fits an integer number of times between the times of occurrence of any pulses), which will be the upper estimate of the transient period. The present period can be an integer number of times less. Our search was carried out for obviously strong pulses ($S/N > 10$ for a total pulse profile over 32 frequency channels). To obtain an estimate of the period, we searched for weaker pulses (up to $S/N = 5$) in the vicinity of the found strong pulses. Weak pulses could not be detected.

The nature of the found RRAT is not clear. It is shown in \citeauthor{Zhou2023}, \citeyear{Zhou2023}, \citeauthor{Tyul'bashev2021}, \citeyear{Tyul'bashev2021} that RRAT is a mixture of known pulsar species. Some of them are pulsars with very long nulling. Some are pulsars with a very wide energy distribution of pulses, and for weak pulsars, strong pulses are observed from the tail of this distribution. Part of the RRAT are pulsars with giant pulses. Two RRATs from the present work have one pulse per 2.5 hours of observations. If we assume that the found RRAT have a period $P = 1$~c, then their nulling will be equal to 99.99\%. That is, we see one pulse out of 10,000. The upper $S_{peak}$ in the average profile of J0249+52 and J0744+55 are 0.3 and 0.2 Jy, respectively. In this case (see Table~1) the observed pulse flux densities exceed the upper estimates of the $S_{peak}$ in the average profile by 40-100 times or more. Such a difference in Speak in a single pulse and in the average profile can be inherent in both pulsars with giant pulses and pulsars with a long tail of pulse energy distribution. The absence of regular radiation speaks in favor of the nulling nature of the found RRAT. However, for an unambiguous choice of the hypothesis about the nature of the found J0249+52 and J0744+55, observations on radio telescopes more sensitive than the LPA LPI are needed.

\section{Discussion of results and conclusion}

The program for monitoring the quality of observations based on data recorded with low time-frequency resolution has shown high efficiency, allowing for rapid response to changes in both the external conditions of observations and internal causes responsible for the deterioration of observations. During the observations, a pulse signal is recorded, visible in one channel or simultaneously in many beams on average every 3 minutes. In general, the quality of observations is high.

The main part of the detected pulses is related to interference, and cluster analysis had to be used to isolate pulsar pulses from the sample. For almost a year of observations, tens of thousands of pulses with signs of "pulsar" were detected in the six-channel data. For each of these events, a pulse at a high frequency arrives earlier than at a low frequency and is registered in one or two adjacent beams. Checking these pulses shows that about half of them belong to pulsars, and the rest of the pulses are various processing artifacts.

In a blind search, 4 known pulsars were found, in which from 1 to more than 10,000 pulses were observed in data with low time-frequency resolution for 7496 observational hours (more than 312 days). In addition to the known pulsars, two new RRATs have been discovered. The appearance of transient pulses on average every 2.5 hours corresponds to known cases (\citeauthor{McLaughlin2006}, \citeyear{McLaughlin2006}). The total number of RRAT discovered in observations at the LPA LPI has reached 48 sources (https://bsa-analytics.prao.ru/en/transients/rrat/).

The analysis showed that, most likely, the transients found are pulsars with very long nulling. For an unambiguous answer, observations on radio telescopes with an instantaneous sensitivity higher than that of the LPA LPI radio telescope are needed.

\section*{Acknowledgements}
The authors are grateful to the antenna team for providing observations at the LPA LPI radio telescope, as well as to L.B. Potapov for help with the preparation of the paper.

\section*{FUNDING}
The study was supported by the Russian Science Foundation (RSF), grant no. 22-12-00236 (https://rscf.ru/project/22-12-00236/).



%

\label{lastpage}
\end{document}